# Evolutionary Dynamics: Unboxing the Wealth of Data Packed in Input-Output Tables

Martin Jaraiz

University of Valladolid

Valladolid, Spain

## Abstract

This paper presents a novel Darwinian Agent-Based Modeling (ABM) methodology for macroeconomic forecasting that leverages evolutionary principles to achieve remarkable computational efficiency and emergent realism. Unlike conventional DSGE and ABM approaches that rely on complex behavioral rules derived from large firm analysis, our framework employs simple "common sense" rules representative of small firms directly serving final consumers. The methodology treats households as the primary drivers of economic dynamics, with firms adapting through market-based natural selection within limited interaction neighborhoods.

We demonstrate that this approach, when constrained by Input-Output table structures, generates realistic economic patterns including wealth distributions, firm size distributions, and sectoral employment patterns without extensive parameter calibration. Using FIGARO Input-Output tables for 46 countries and focusing on Austria as a case study, we show that the model reproduces empirical regularities while maintaining computational efficiency on standard laptops rather than requiring supercomputing clusters.

Key findings include: (1) emergence of realistic firm and employment distributions from minimal behavioral assumptions, (2) accurate reproduction of the initial Social Accounting Matrix values through evolutionary dynamics, (3) successful calibration using only 5-6 country-specific parameters to complement the FIGARO data, and (4) computational performance enabling full simulations on consumer hardware. These results suggest that evolutionary ABM approaches can provide robust policy insights by capturing decentralized market adaptations while avoiding the computational complexity of traditional DSGE and comprehensive ABM models.

## 1. Introduction

Agent-Based Models (ABMs) have emerged as powerful tools for understanding complex economic systems, offering advantages over traditional equilibrium models by capturing heterogeneity, non-linear interactions, and emergent phenomena (Tesfatsion, 2006; Farmer & Foley, 2009). However, the computational complexity of comprehensive ABMs, which attempt to model detailed behavioral rules for all agent types, has limited their practical application in policy analysis and forecasting (Poledna et al., 2023).

In a recent comprehensive survey, Dawid and Delli Gatti (2018) identify seven main families of macroeconomic ABMs and provide a systematic comparison of their structures and modeling assumptions. They note that despite the diversity of approaches, there is an emerging common core of agent-based macroeconomic modeling, characterized by heterogeneous agents with bounded rationality, decentralized interactions, and emergent aggregate dynamics. This survey provides crucial context for understanding how the Darwinian approach presented here relates to and differs from existing ABM frameworks.

This paper introduces a fundamentally different approach, Figure 1: Darwinian Agent-Based Modeling (DABM), which applies evolutionary principles to achieve both computational efficiency and empirical validity. Rather than specifying complex behavioral rules ex-ante, our framework allows realistic economic patterns to emerge through market-based selection mechanisms operating on agents following simple heuristics.

It is important to emphasize that the Darwinian framework presented here establishes a foundational architecture that deliberately prioritizes computational efficiency and emergent behavior over predetermined complexity. While the current implementation demonstrates remarkable success with minimal behavioral specifications, the modular nature of the approach allows for systematic incorporation of additional complexity schemes in subsequent developments. These enhancements can include sector-specific behavioral rules, financial market microstructure, international trade dynamics, and environmental constraints, each adding layers of realistic detail and accuracy to the model's predictions. However, the key innovation and distinguishing feature of the Darwinian approach lies not in the potential for adding complexity, but in its fundamental ability to deploy a functioning economic system from scratch using only evolutionary principles and Input-Output constraints, generating empirically valid patterns without requiring extensive a priori behavioral assumptions.

Moreover, the fundamental architecture of the Darwinian method provides researchers with unprecedented flexibility in exploring different behavioral assumptions. Rather than being constrained by the minimal specifications demonstrated in this paper, researchers can introduce sophisticated behavioral rules from the outset - including complex expectation formation mechanisms, detailed market analysis capabilities, strategic planning algorithms, or any other behavioral models drawn from economic theory or empirical observation. The evolutionary framework will then reveal how economic systems emerge and self-organize under these alternative behavioral specifications. This capability transforms the Darwinian approach into a powerful experimental platform for testing competing theories of economic behavior, as the method's validity does not depend on any particular set of behavioral assumptions but rather on the evolutionary selection process that determines which behaviors survive and proliferate in the market environment.

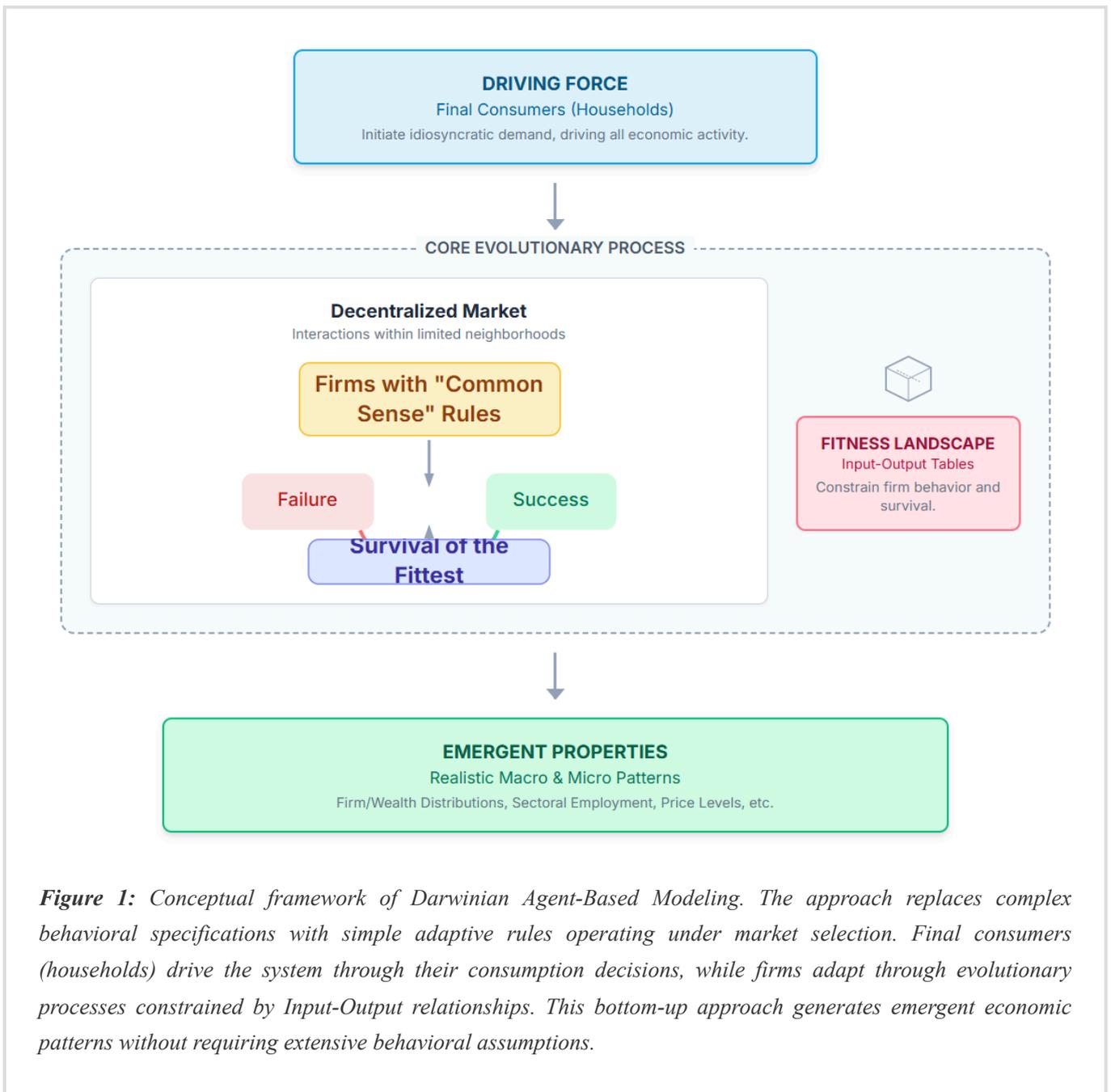

*Figure 1:* Conceptual framework of Darwinian Agent-Based Modeling. The approach replaces complex behavioral specifications with simple adaptive rules operating under market selection. Final consumers (households) drive the system through their consumption decisions, while firms adapt through evolutionary processes constrained by Input-Output relationships. This bottom-up approach generates emergent economic patterns without requiring extensive behavioral assumptions.

The motivation for this approach stems from three key observations:

First, most economic activity involves small firms that lack the resources for sophisticated market analysis. These firms operate on simple rules of thumb, adjusting prices based on immediate market feedback and maintaining inventory proportional to recent demand. Yet collectively, these simple behaviors generate the complex patterns observed in real economies.

Second, final consumer demand represents the ultimate driver of economic activity. By modeling the economy from the perspective of household consumption decisions propagating through production networks, we align the model's causal structure with empirical reality.

Third, Input-Output tables encode rich information about economic structure that can serve as fitness constraints for evolutionary dynamics. Rather than calibrating numerous behavioral parameters, we allow the I-O structure to guide the selection of viable firm behaviors and market configurations.

## 2. Methodology

### 2.1 Model Architecture

The DABM framework presented here consists of four primary agent types operating within locally constrained interaction neighborhoods, Figure 2. Each agent type plays a specific role in the economic system, with behaviors designed to capture essential features while maintaining computational simplicity.

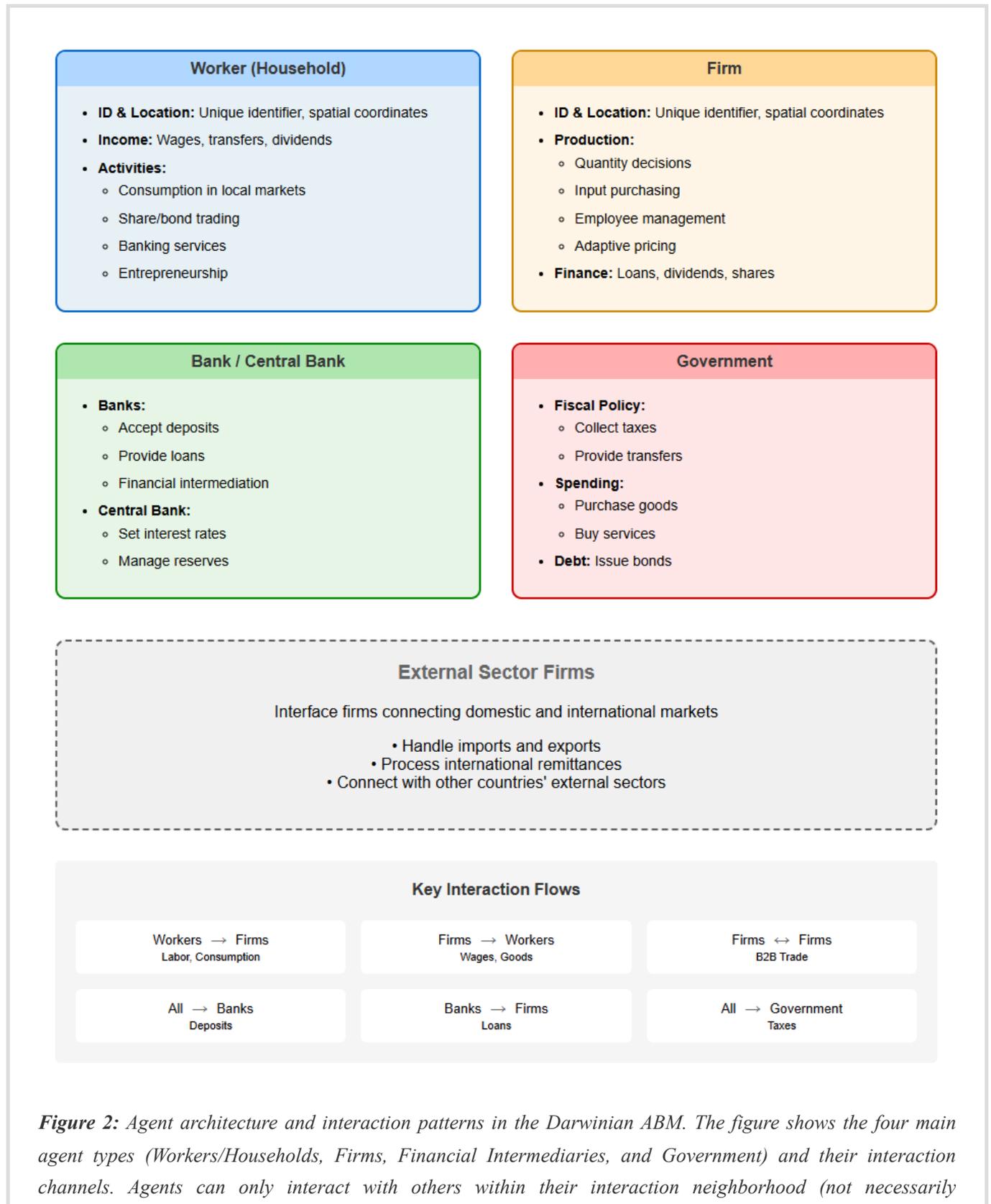

*Figure 2:* Agent architecture and interaction patterns in the Darwinian ABM. The figure shows the four main agent types (Workers/Households, Firms, Financial Intermediaries, and Government) and their interaction channels. Agents can only interact with others within their interaction neighborhood (not necessarily

> geographic), creating realistic market segmentation and specific competition patterns . The External Sectors are firms operating at different international trade interfaces. The decentralized structure allows complex patterns to emerge from neighborhood interactions.

**Workers (Households)**

Each worker agent possesses a unique identifier and location defining their interaction radius. Income derives from wages, transfers, and firm ownership dividends. Workers make consumption decisions within local neighborhoods, can start new firms based on entrepreneurial probabilities, and manage portfolios of shares and bonds. This simple structure captures the essential features of household economic behavior without requiring utility maximization or complex planning algorithms.

**Firms**

Production entities are characterized by location determining supplier and customer networks. Each firm operates according to sector-specific production functions derived from I-O tables. Pricing adapts through local market feedback - successful sales lead to price increases while failed attempts trigger reductions. Employment and inventory management respond dynamically to demand, with survival dependent on maintaining positive cash flow. This design allows firms to discover viable strategies through market interaction rather than optimization.

**Financial Intermediaries**

Banks and the central bank manage deposit and loan services, with the central bank setting interest rates and managing reserves. The simplified financial sector captures essential monetary transmission mechanisms without modeling complex financial instruments or strategies.

**Government**

The fiscal authority collects taxes, provides transfers, procures goods and services, and issues bonds for deficit financing. Government behavior follows simple rules based on target spending levels and automatic stabilizers, avoiding the need to model political economy considerations.

**External Sector**

Interface firms manage import/export transactions, international remittances, and cross-border supply chains. This abstraction allows the model to capture international trade effects without explicitly modeling foreign economies.

## 2.2 Behavioral Rules

The model operates on five core behavioral principles designed to capture essential market dynamics while maintaining computational tractability:

**Adaptive Pricing:** After each transaction attempt, both buyer and seller adjust price expectations by a small calibrated factor, p = p + k*p, e.g. universal constant k=0.002 . This bilateral adjustment mechanism allows prices to emerge from decentralized negotiations, discovering market-clearing levels without centralized auctioneers or equilibrium assumptions.

**Wealth-Based Consumption:** Household expenditure follows a country-specific calibrated propensity to consume applied to total wealth, subject to liquidity constraints. This rule captures both wealth effects and precautionary savings without requiring intertemporal optimization.

**Entrepreneurial Entry:** New firm creation follows country-specific probabilities calibrated to match observed firm birth rates. Workers sensing unmet demand in their neighborhood probabilistically establish new firms, with sector selection influenced by local consumption patterns.

**Demand-Driven Production:** Firms maintain inventory at a multiple of previous period demand. This simple rule ensures market liquidity while avoiding excessive stocks, capturing essential inventory behavior without sophisticated forecasting.

**Local Interaction:** All transactions occur within defined neighborhoods, where "neighborhood" refers not exclusively to geographic proximity but rather to the empirical observation that each agent maintains interactions with a relatively stable and limited set of other agents over time. This constraint captures the reality that economic relationships tend to persist through repeated interactions, whether due to spatial proximity, established business relationships, social networks, or sectoral specialization. Agents do not randomly interact with the entire population but instead develop and maintain connections with specific subsets of other agents, creating realistic patterns of market segmentation and localized competition. This interaction structure allows for the emergence of heterogeneous market conditions and price disparities across different parts of the economy, reflecting the imperfect information and relationship-based nature of real economic activity.

## 2.3 Evolutionary Dynamics

The model's evolutionary character emerges through several interconnected mechanisms that allow economic structure to develop endogenously rather than being imposed by modeler assumptions.

**Market Selection** operates continuously as firms unable to maintain positive cash flow exit the market, while successful firms expand employment and production. This selection process requires no centralized optimization, mimicking real market dynamics where survival depends on efficiently serving customer needs.

**Emergent Specialization** develops as firms naturally find sectoral niches based on local demand patterns and I-O constraints. Starting from homogeneous initial conditions, the economy develops a complex division of labor through trial and error rather than planning.

**Price Discovery** occurs through decentralized bilateral negotiations, with successful price points propagating through local networks. The evolutionary process discovers prices that balance supply and demand without requiring tatonnement or market clearing assumptions.

**Structural Evolution** allows the economy's sectoral composition, firm size distribution, and employment patterns to emerge endogenously. This approach lets the model discover realistic economic structures through the same processes that generate them in real economies.

## 2.4 Input-Output Integration

The integration of Input-Output tables provides crucial structure to the evolutionary process. Rather than treating I-O coefficients as mere calibration targets, they act as fitness constraints that shape which firm behaviors can survive in the market.

*Figure 3: Integration of Input-Output tables in the evolutionary framework. The Social Accounting Matrix (SAM) provides the fitness landscape for firm evolution. When firms in the AgroPesc sector generate revenue, they must allocate it according to the proportions shown: wages (€4,259 per €48,021 revenue), intermediate consumption to other sectors, and value added. Firms violating these proportions face competitive disadvantage and eventual market exit. This mechanism ensures that emergent economic structures align with empirical I-O relationships.*

When a firm (Figure 3) from a particular sector generates revenue, it must purchase inputs and pay wages according to the ratios specified in the I-O table. Firms that deviate significantly from these proportions face higher costs or insufficient inputs, leading to competitive disadvantage. This mechanism ensures that the emergent economic structure aligns with empirical intersectoral relationships without requiring explicit enforcement.

## 3. Implementation Framework

### 3.1 Initialization and Deployment

The model deployment, Figure 4, follows a unique approach where the economic system begins without firms, allowing market structures to emerge entirely through evolutionary processes. This strategy ensures that economic patterns arise from market dynamics rather than modeler assumptions.

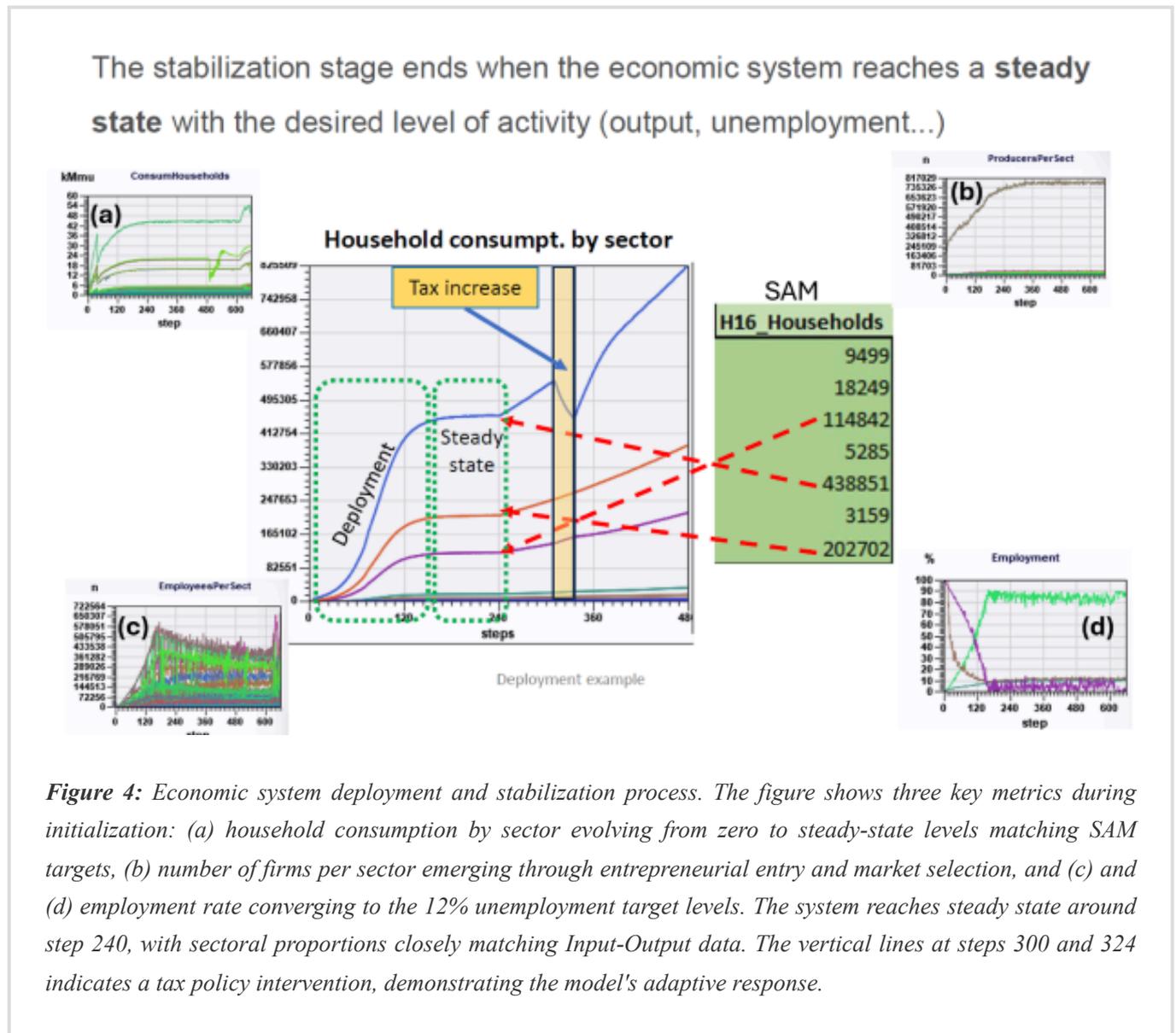

*Figure 4:* Economic system deployment and stabilization process. The figure shows three key metrics during initialization: (a) household consumption by sector evolving from zero to steady-state levels matching SAM targets, (b) number of firms per sector emerging through entrepreneurial entry and market selection, and (c) and (d) employment rate converging to the 12% unemployment target levels. The system reaches steady state around step 240, with sectoral proportions closely matching Input-Output data. The vertical lines at steps 300 and 324 indicates a tax policy intervention, demonstrating the model's adaptive response.

**Phase 1: Agent Creation.** Workers are instantiated based on the target simulation size, typically using a scaling factor of 10 workers per million population per sector. For the Austrian case study with 3,200 simulated workers, this represents a 1:1,562 scaling of the actual workforce.

**Phase 2: Demand Activation.** Households begin consuming according to their wealth and propensity to consume, initially attempting purchases that fail due to the absence of suppliers. These failed attempts create market signals for profitable opportunities.

**Phase 3: Firm Emergence.** Workers probabilistically establish firms in response to unmet demand, with sector selection influenced by consumption patterns. New firms inherit production functions from their sector's I-O coefficients.

**Phase 4: Market Stabilization.** Through repeated interactions, the system evolves toward quasi-steady state with balanced firm demographics and sectoral alignment with I-O structures.

## 3.2 Technical Architecture

The model, implemented in C++ (20k code lines), is still under development, with careful attention to computational efficiency and self-calibration algorithms. Memory management uses contiguous layouts for cache efficiency, with spatial indexing accelerating neighborhood queries. While currently using single-thread, the architecture supports parallel execution through spatial decomposition.

## 3.3 Calibrated Parameters and Emergent Properties Categories

There are two results data categories: calibrated and emergent

Example:

Households

- **We calibrate** *k* to achieve a good fit to the households **Consumption** SAM values:

$$C^\star_{h,t} = I^{Mean}_{h,t} + \kappa \cdot \left(W_{h,t} - \Phi \cdot I^{Mean}_{h,t}\right)$$

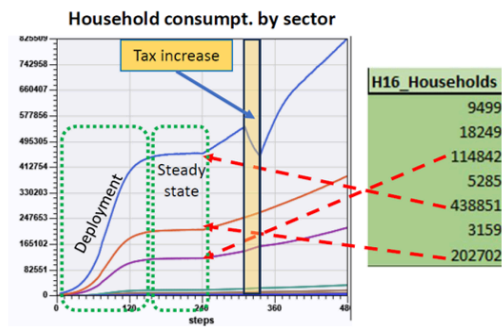

- **We observe** the evolutionary **emergence** of a non-predefined **Wealth Distribution**

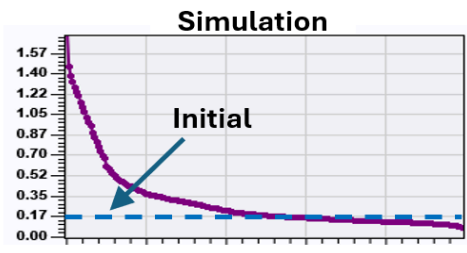
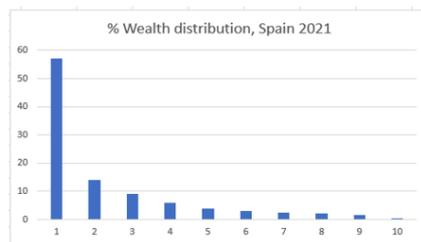

*Figure 5: Calibrated Parameters and Emergent Properties Categories.*

Figure 5 illustrates a fundamental distinction in the Darwinian Agent-Based Model between calibrated parameters and emergent properties. The upper panel demonstrates the calibration process for household consumption by sector, where the parameter κ (kappa) is adjusted to ensure the model reproduces the Social Accounting Matrix consumption values accurately. This calibration represents a necessary alignment with empirical input-output data that constrains the model's behavior.

In contrast, the lower panel showcases one of the model's most compelling validation results: the emergence of realistic wealth inequality without any explicit distributional assumptions. Starting from an equal wealth distribution across all agents, the model generates a wealth concentration pattern that closely matches empirical data from Spain in 2021, with the top decile controlling approximately 60% of total wealth. This emergent property arises solely from the interaction of simple behavioral rules, market selection processes, and Input-Output constraints, demonstrating that the model captures fundamental wealth accumulation dynamics without requiring pre-specification of inequality-generating mechanisms.

This distinction between calibrated inputs and emergent outputs provides crucial validation for the Darwinian approach, showing that minimal calibration requirements can yield rich, empirically accurate emergent patterns through evolutionary market processes.

## 4. Calibration Methodology

### 4.1 Minimal Parameter Set

Unlike conventional ABMs requiring hundreds of parameters, the DABM approach needs only 5-6 country-specific values. This dramatic reduction in parameter space stems from allowing market evolution to determine behavioral patterns rather than specifying them exogenously.

Table 1: Country-specific parameters for Austrian model calibration

| Parameter | Value | Description |
| --- | --- | --- |
| Active_Population | $5.0 \times 10^6$ | Total workforce |
| Unemployment_Rate | 4.4% | Target jobless rate |
| Households_Wealth_Target | €$1.5 \times 10^{12}$ | Aggregate household wealth |
| Propensity_To_Consume | 0.55 | Marginal consumption rate |
| Firms_Birth_Per_Year | 36,000 | Annual firm creation rate |
| Producer_Startup_Probability | 0.007 | Calibrated entrepreneurship rate |

### 4.2 Evolutionary Calibration Process

The calibration proceeds through evolutionary adaptation rather than optimization. Starting with initial parameter guesses, the system evolves toward configurations that match empirical targets. If firm creation rates deviate from observed values, the startup probability adjusts incrementally. The evolutionary dynamics naturally correct sector-specific imbalances without requiring manual intervention.

Convergence monitoring uses multiple indicators including sectoral output alignment with I-O totals, employment distribution across sectors, firm size distribution, and aggregate macroeconomic ratios. The system demonstrates remarkable stability once calibrated, maintaining steady-state behavior without parameter drift.

A key computational advantage of the Darwinian approach is that the evolutionary dynamics and market selection mechanisms produce stable emergent patterns in single simulation runs of a few thousand workers. In contrast, comprehensive ABMs typically require averaging across 500 or more Monte Carlo simulations of the entire workforce to achieve statistically reliable results, necessitating supercomputing resources.

# 5. Experimental Results

## 5.1 Austrian Case Study Overview

To demonstrate the effectiveness of the Darwinian Agent-Based Modeling approach, we present detailed results from simulating the Austrian economy using 2010 FIGARO Input-Output data. The simulation scale balances computational tractability with sufficient agent diversity to capture emergent phenomena.

## 5.2 Calibration: Convergence to desired Social Accounting Matrix

One of the most striking results is the model's ability to reproduce Social Accounting Matrix values through purely emergent processes. Starting from an economy with zero firms, Figure 6, the evolutionary dynamics generate sectoral structures closely matching empirical I-O data.

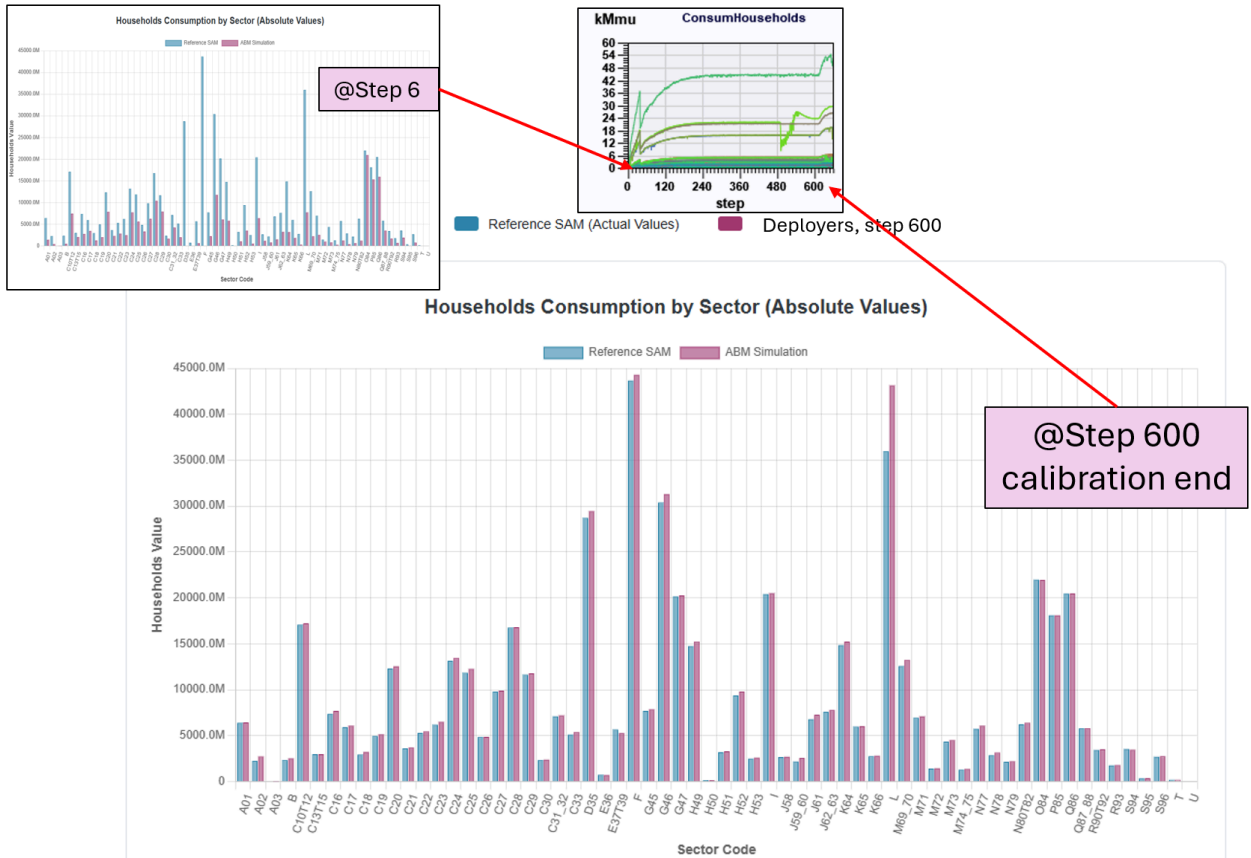

*Figure 6:* Convergence to Social Accounting Matrix values after 600 months of evolution. The blue bars show reference SAM values from FIGARO I-O tables, while red bars display emergent values from the simulation. Despite starting with zero firms and no predetermined structure, the evolutionary process generates sectoral consumption patterns closely matching empirical data. Major sectors achieve accuracy within 5%, validating the effectiveness of I-O constraints as evolutionary fitness criteria.

### 5.3 Emergent Firm and Employment Distributions

The model generates realistic firm size distributions without any explicit size-dependent behavioral rules, Figure 7. This emergence occurs through market selection favoring efficient scales, local competition limiting concentration, and sectoral differences in capital intensity driving size variation.

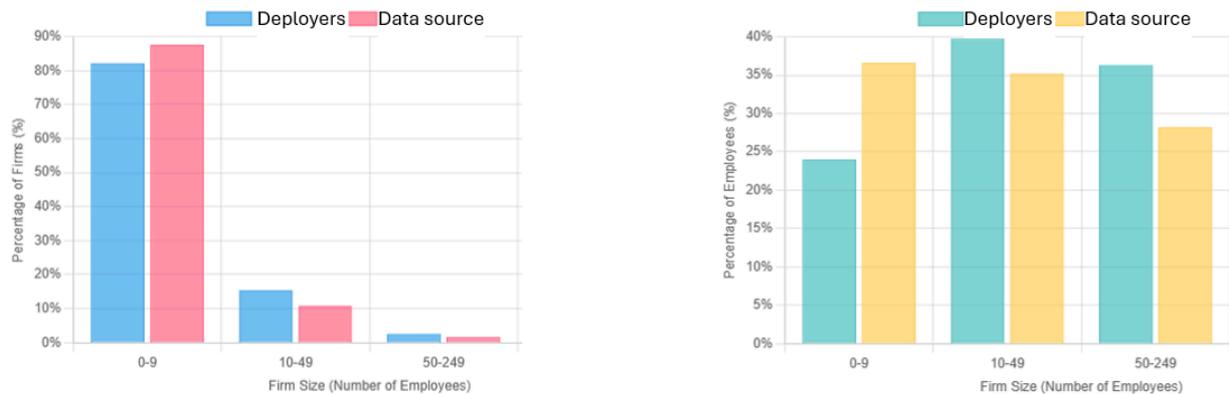

*Figure 7:* Emergent firm and workforce distributions compared with empirical data. Left histogram displays firm size distributions, where both model and data show approximately 87.6% micro firms (0-9 employees), 10.8% small firms (10-49), 1.6% medium firms (50-249). The right panel reveals that employment concentrates in medium and large firms despite their small numbers. These distributions emerge without explicit size-dependent rules, arising solely from market selection processes.

## 5.4 Wealth Distribution Evolution

One of the most compelling validations of the model's market mechanisms is the emergence of realistic wealth inequality from initially equal distributions, as shown in Figure 6 for a six sectors SAM of Spain (2008). This process operates through multiple channels including differential wages and dividends from successful firms, entrepreneurial wealth accumulation, compound effects of initial random variations, and local clustering creating regional disparities.

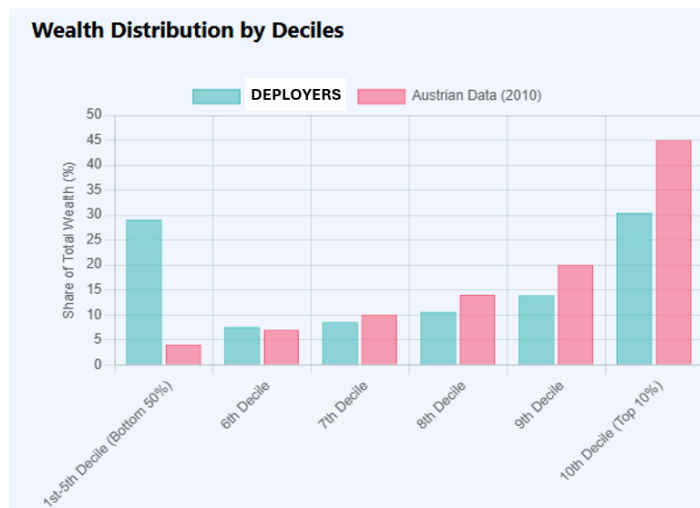
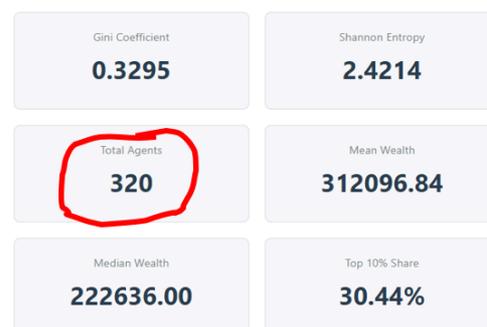
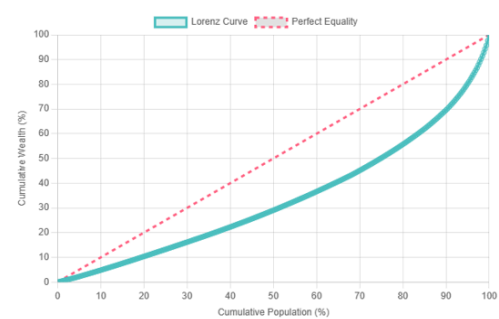

*Figure 8:* Emergent wealth distribution. These distributions emerge without explicit size-dependent rules, arising solely from market selection processes. The figure shows the result from a very small simulation of only 320 workers, for Austria (2010) with the 64-sectors FIGARO IO table.

### 5.5 Dynamic Response Analysis

To probe the model's behavior beyond steady-state properties, we conducted preliminary policy intervention experiments. These dynamic experiments were performed on a six sectors SAM and are shown in Figure 6 (using the previous version of the program, because the version currently under development is still operating at constant prices and other constraints, to focus on the maximization of the accuracy of the production deployment stage). The model's response to a tax increase demonstrates realistic dynamics including immediate consumption decline, firm exit and employment adjustment, and gradual convergence to new equilibrium.

### 5.6 Computational Performance

The Darwinian approach achieves remarkable computational efficiency compared to traditional DSGE and comprehensive ABMs, Figure 9. For the Austrian economy simulation on a standard laptop (Intel Core i7, 8GB RAM), each simulation month processes in approximately 9 seconds, full 600-month calibration completes in 90 minutes, and subsequent 12-month forecast runs require only 2 minutes.

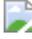

*Figure 9:* *Computational performance of the Darwinian ABM implementation. Panels (a), (b), (c): The model simulates 3,200 workers and approximately 400 firms on a single CPU core of a standard Windows laptop. Calibration over 600 months requires a 90 minutes single run, while production runs of 12 months complete in 2 minutes. This represents many orders of magnitude efficiency gain compared to DSGE and comprehensive ABMs, panel (d), requiring high-performance computing clusters for similar-scale simulations. Panel (c) shows the annualized monthly birth and death of firms as an alive process around a quasi-constant average.*

## 6. Discussion

### 6.1 Theoretical Implications

The success of the Darwinian Agent-Based Modeling approach in reproducing complex economic patterns from simple behavioral rules has profound implications for economic theory. It demonstrates that much of the observed complexity in economic systems may emerge from simple interactions shaped by market selection rather than from sophisticated individual optimization.

This finding extends Simon's bounded rationality concept by showing how market evolution generates aggregate patterns indistinguishable from those produced by optimizing agents. The evolutionary process effectively solves coordination problems that individual agents cannot solve through planning, suggesting a fundamental revision in how we conceptualize micro-macro relationships.

The role of Input-Output tables as fitness landscapes rather than mere accounting relationships reveals their deeper significance. I-O structures encode fundamental constraints on viable economic organization, discoverable through market processes. This interpretation provides new theoretical foundations for understanding why certain economic structures persist across different institutional contexts.

### 6.2 Methodological Advantages

As Dawid and Delli Gatti (2018) note in their comprehensive survey of agent-based macroeconomic models, the field has developed multiple distinct approaches, each with different behavioral specifications and computational requirements. The Darwinian approach presented here represents a significant departure from these existing frameworks by demonstrating that complex economic patterns can emerge from much simpler behavioral rules when combined with market selection mechanisms, thereby achieving the computational efficiency necessary for practical policy applications.

The Darwinian approach addresses key criticisms of agent-based modeling while maintaining its advantages. The dramatic reduction in free parameters - from hundreds to just 5-6 - eliminates concerns about overfitting while maintaining empirical validity. Evolutionary calibration replaces sophisticated optimization algorithms with market-based discovery of viable behaviors. Key advantages summary:

**Emergent Realism:** Simple behavioral rules + simulation of market activity in everyday life suffice to generate realistic economic patterns without complex prior specifications.

**Leveraging data:** The emergence of these realistic patterns is made possible by the wealth of information about the economic system conveyed by the I-O tables, used as constraints for the fitness of firms and agents' behaviors.

**Policy robustness:** The models capture creative and decentralized market adaptations, providing more reliable insights than fixed behavioral frameworks.

**Discovery capability:** Evolutionary dynamics reveal unexpected phenomena beyond modelers' assumptions, crucial for crises and transitions.

**Calibration parsimony:** Natural selection reproduces empirical regularities without large parameter adjustments or risks of overfitting.

**Computational efficiency:** The Darwinian approach demonstrates remarkable computational efficiency by producing economically meaningful results from single simulation runs on standard laptops. Traditional comprehensive ABMs, by comparison, require averaging across hundreds of Monte Carlo simulations of 1000x the number of simulated workers on high-performance computing clusters to achieve comparable stability and reliability.

## 6.3 Future Extensions

The framework's agent-based foundation provides natural extension points for incorporating additional complexity while maintaining computational efficiency. Future developments could interface the core economic model with additional systems including ecological constraints through the System of Environmental Economic Accounting (SEEA), demographic or social network dynamics, and epidemiological models. These extensions leverage the agent-based architecture while maintaining the computational efficiency of the evolutionary approach, enabling integrated analysis of complex socio-economic-environmental systems.

## 6.4 Policy Analysis Applications

The framework's efficiency and robustness make it ideal for policy analysis. Real-time adaptation allows agents to respond naturally to policy changes without predetermined elasticities. Heterogeneous impacts emerge from agent positions in the market structure rather than assumed categories. Unintended consequences become visible through market evolution, revealing effects missed by partial equilibrium analysis. Some preliminary policy experiments demonstrate the framework's capabilities across multiple domains. Fiscal policy analysis reveals how tax changes propagate through employment and demand channels.

Central banks and international organizations face critical choices about modeling priorities. Investing in machine learning integration could dramatically improve forecasting accuracy but risks creating black-box models unsuitable for policy communication. Developing comprehensive heterogeneous agent models could better capture distributional concerns but requires computational resources that strain operational timelines. Incorporating climate dynamics addresses existential risks but complicates models already pushing tractability limits. The framework presented here drastically reduces the computational resources required, while its agent-based foundation circumvents model tractability obstacles. In combination with real-time data centers this scheme could provide nowcasting capabilities complementary to current black-box machine learning models.

## 7. Conclusions

This paper has introduced and validated a Darwinian approach to Agent-Based Economic Modeling that achieves remarkable efficiency while maintaining empirical validity. By replacing complex behavioral specifications with simple adaptive rules operating under market selection, realistic economic patterns emerge naturally from the interaction of boundedly rational agents within Input-Output constrained environments.

A key innovation of our approach is that the model's evolutionary dynamics create self-organizing stability that eliminates the need for Monte Carlo averaging during regular operation, leveraging I-O tables as fitness landscapes encoding fundamental economic constraints. This yields many orders of magnitude higher computational efficiency, generating empirically valid distributions without explicit mechanisms, and maintaining robustness through market-based adaptation.

Our results suggest that much economic complexity arises from simple interactions shaped by market selection rather than sophisticated optimization, an insight with profound implications for theory, empirical analysis, and policy design. The computational efficiency and minimal parameterization of the Darwinian Agent-Based Modeling approach make it particularly suitable for practical applications requiring rapid analysis based on readily available input data like I-O tables, while future development promises to extend capabilities without sacrificing the core advantages of evolutionary simplicity.

As economic challenges grow increasingly complex and interconnected, tools that combine empirical validity with practical usability become essential, and this approach offers a path toward economic models that are simultaneously more realistic, more tractable, and more scientifically parsimonious than current alternatives.

The future of economic modeling lies in combining the empirical rigor of Input-Output analysis with the dynamic richness of Darwinian Agent-Based Modeling. Starting from dominant mechanisms provides a powerful new technology for understanding economic systems in all their complexity, while maintaining accessibility on standard computing hardware. This democratization of sophisticated economic modeling opens new possibilities for research and policy analysis.

**Data Availability**

The FIGARO Input-Output tables used in this study are publicly available from Eurostat. The simulation software "Deployers" will be made available upon publication.